# Enhanced sensing performance by the plasmonic analogue of electromagnetically induced transparency in active metamaterials

Zheng-Gao Dong,[1,*] Hui Liu,[2] Jing-Xiao Cao,[2] Tao Li,[2] Shu-Ming Wang,[2] Shi-Ning Zhu,[2] and X. Zhang[3]

[1]*Physics Department, Southeast University, Nanjing 211189, China*

[2]*National Laboratory of Solid State Microstructures, Nanjing University, Nanjing 210093, China*

[3]*5130 Etcheverry Hall, Nanoscale Science and Engineering Center, University of California, Berkeley, California 94720-1740, USA*

The gain-assisted plasmonic analogue of electromagnetically induced transparency (EIT) in a metallic metamaterial is investigated for the purpose to enhance the sensing performance of the EIT-like plasmonic structure. The structure is composed of three bars in one unit, two of which are parallel to each other (dark quadrupolar element) but perpendicular to the third bar (bright dipolar element), The results show that, in addition to the high sensitivity to the refractive-index fluctuation of the surrounding medium, the figure of merit for such active EIT-like metamaterials can be greatly enhanced, which is attributed to the amplified narrow transparency peak.

Plasmonically induced transparency, also called plasmonic analogue of the electromagnetically induced transparency (EIT),[1-4] is an interesting phenomenon in metamaterials wherein bright-mode energy is transformed into a dark mode through plasmon coupling between the adjacent resonant elements. This phenomenon results in a narrow transmission peak with a high quality factor within the originally broad stop band (bright mode) of the transmission spectrum. Here, the bright mode indicates a far-field excitable resonance with radiative losses, such as the dipolar resonance of a single-rod array. The dark mode, on the contrary, cannot be excited by a far-field source and does not radiate into outer environment (high quality factor). For example, the quadrupolar resonance of a rod-pairing array is "dark" to incident waves when the polarized *H* field is parallel to the rod-pair plane. Such a metamaterial version with EIT feature is considered as a plasmonic analogue of the EIT phenomenon in atomic physics. From the quantum point of view,[5,6] it happens when the excitation pathway |0>-|1> interferes destructively with another pathway |0>-|1>-|2>-|1>, where |0>-|1> means a dipole-allowed transition (bright mode) from the ground state |0> to the excitation state |1>, whereas direct transition |0>-|2> is forbidden (dark mode). Nevertheless, the excitation state |2> can be accessed through plasmon-coupling transition |1>-|2>.

Generally, although it is similar regarding the capability of opening a narrow transparency window within a transmission stop band, the plasmonically induced transparency differs from a native EIT in atomic system for several aspects. First, it happens through a high efficiency of plasmon coupling between the bright and dark modes, rather than by an external beam to control the EIT window. Second, the EIT-like feature occurs only when the dark mode is at the identical resonant frequency range with the bright mode. Third, the EIT-like phenomenon can only suppress the radiative damping by transferring the energy from the bright (radiative) mode into the dark (subradiant) mode. In other words, the intrinsic ohmic loss inside the metal can never be compensated in these EIT-like structures. Consequently, although EIT-like metamaterials integrate the dual advantages of EIT-like narrow window (dark-mode excitation) and surface plasmon resonance (strong field confinement) for sensor applications,[7,8] low transmittance due to the intrinsic ohmic loss in EIT-like metamaterials needs to be improved for a better sensing efficiency. Recently, nonlinear active media such as quantum dots have been used to compensate for the ohmic loss in plasmonic structures.[9] In particular, stimulated emission of radiation can amplify the surface plasmon resonance considerably in such active plasmonic systems,[10] inspiring proposals on interesting nonlinear optical devices, such as nanolaser or lasing spaser.[11,12] In this work, we demonstrate that the resonance amplification in active plasmonic metamaterials can be applied to EIT-like metamaterial sensors for a great enhancement of the sensing capability.

For the EIT-like phenomenon in metallic metamaterials, the bright element is either a single bar or a split ring resonator, if only it is directly excitable by the incident waves and meanwhile has a broader linewidth than does the dark element, usually comprising split ring resonator,[13-16] pairing bars,[2-4,17] and double rings.[18,19] In this work, we take the previously studied three-bar configuration[2-4] to investigate the *active* plasmonic analogue of EIT in order to enhance the performance of sensing the refractive-index fluctuation of the surrounding medium.

Figure 1 shows a schematic illustration of the unit cell of the metallic metamaterial with structural parameters as follows: $a = 460$ nm, $b = 80$ nm, $t = 40$ nm, $g = 300$ nm, $h = 110$ nm, and $l = 260$ nm. The periodicity in both $x$ and $y$ direction is $p_x = p_y = 700$ nm, while only single unit layer is considered in the $z$ direction. Under the incident polarization situation in Fig. 1(a), the dipolar oscillation is a bright mode in the perpendicularly stacked bar, whereas the quadrupolar mode is a dark mode inherent to the double parallel bars which can only be excited through plasmon coupling by introducing the symmetry-broken shift $s$ [Fig. 1(b)]. A full-wave finite element method is used for the simulation,[12,18] which automatically takes account of the diffraction effect by solving the Maxwell's Equations with given material property, structure, and boundary conditions. The perfect electric and magnetic boundaries are used in compliance with the incident configuration in Fig. 1(a). The metal is silver with Drude dispersion ($\omega_p = 1.37 \times 10^{16} \text{s}^{-1}$ and $\gamma = 8.5 \times 10^{13} \text{s}^{-1}$),[20] for which the substrate is glass with an index of refraction of 1.55. The metallic structure is embedded in a host material of polymethyl methacrylate (PMMA, index of refraction 1.49), in which the gain medium (for example, PbS semiconductor quantum dots) is sparsely doped. A frequency-dependent complex permittivity is used to characterize the active property with gain coefficient $\alpha = (2\pi/\lambda)\text{Im}(\sqrt{\varepsilon' + i\varepsilon''})$, where $\varepsilon'$ and $\varepsilon''$ are the real and imaginary parts of the permittivity for the active system, respectively.[12,21]

To evaluate the EIT-like response of this structure in the absence of gain, the transmittance spectra for $\alpha = 0\,\text{cm}^{-1}$ at different asymmetry shifts *s* are presented in Fig. 2. It is found that more asymmetry degree by increasing *s* results in broader EIT range as well as larger transmittance, corresponding to the so-called Autler-Townes doublet in atomic systems.[22,23] This broadening, although with enough transmission intensity, is not preferable for sensing performance because of the degenerated figure of merit (FOM).[24-27] On the other hand, a smaller asymmetry shift *s* results in a more pronounced narrow EIT feature, but unpleasantly the transmittance is suppressed.[3] Therefore, a narrow transmission peak with high transmittance for sensing purpose could not be managed simultaneously in a passive EIT-like metamaterial, unless an active medium will be introduced to compensate for the intrinsic ohmic loss.[2]

Figure 3 shows the gain-assisted transmittance transition of the EIT-like window at $s = 20\,\text{nm}$. It is found that the transparency peak enhances its transmittance dramatically, even to a level with two orders of intensity amplification for the gain coefficient of $850\,\text{cm}^{-1}$. Meanwhile, it always keeps the narrow characteristic of an EIT window regardless of the gain value. The full width at half maximum (FWHM), originally 2.5 THz (about 19.5 nm) without gain assistance, reaches the maximum value of 0.3 THz (about 2.0 nm). This greatly enhanced sharpness of the transparency peak is valuable for sensing applications. Experimentally, a typical gain coefficient value about 300 cm$^{-1}$ at room temperature for PbS quantum dots was reported.[28] In reality, the gain coefficient depends on the PbS doping density, fabrication procedure, power of the pumping source, and operation temperature, etc..[29] According to Fig. 3, a gain coefficient value of 200~400 $\text{cm}^{-1}$ is sufficient to reach a greatly improved transparency window for sensing application. Moreover, if the gain coefficient reaches the value around 600~850 $\text{cm}^{-1}$, the best sensing ability can be obtained. Anyway, a realistic gain value in hundreds of $\text{cm}^{-1}$ is sufficient for a large enhancement of the transparency window, partially attributed to the underlying EIT-like characteristic that minimizes resonant losses. In addition, the nonmonotonic dependence of the transmittance enhancement on the gain coefficient has been explained elsewhere.[12]

To investigate the sensing performance of the gain-assisted EIT-like transparency window, calculated results about sensitivity and FOM are presented in Fig. 4. According to the formula $FOM = m\,(\text{nm}\,\text{RIU}^{-1})\,/\,FWHM\,(\text{nm})$, the sensitivity $m$ is the linear slope of the resonance wavelength shift over the refractive-index change, while FOM takes a further consideration of the resonant lineshape.[26,27] Above all, the EIT-like transparency window exhibits a distinct resonance shift with respect to a small fluctuation in the refractive index of the surrounding medium, even for $\Delta n = 0.01$ [Fig. 4(a), for $\alpha = 850\,\text{cm}^{-1}$]. Therefore, it offers an excellent potential for ultrahigh resolution situations, such as biosensing and gas detection. The wavelength shift of the transparency peak per refractive-index unit shows a high sensitivity about 680 nm/RIU. In addition, the sensing performance in terms of FOM, by taking into account the sharpness of a transmission peak, offers another advantage due to the assistance of the gain medium. As is shown in Fig. 4(b), the gain dependence of the FOM actually indicates a great improvement in the sensing performance, from which the gain-assisted FOM can reach a maximum value as large as 400 when a suitable gain coefficient is introduced. However, it should be noticed that the FOM could decrease substantially when the refractive index of the surrounding medium changes widely so that it may result in a shifted EIT-like transparency frequency far from the original gain value of an active medium, which is frequency dependent and generally has a Gaussian distribution. In this sense, the gain-assisted FOM also depends, indirectly, on the refractive index of surrounding medium. Therefore, an *in situ* control of the gain coefficient should be desired to keep the enhanced FOM as is throughout a wide refractive-index change of the surrounding medium.[29]

In summary, a sensing scheme that incorporates the active mechanism in a plasmonic metamaterial and the narrow transparency feature of the EIT phenomenon is proposed to enhance the sensing performance. Three distinguishing characteristics can be concluded. First, the EIT-like resonance, originating from the dark-mode excitation by plasmon coupling, has a great sensitivity as much as 680 nm/RIU. Second, the gain-assisted FOM in the EIT-like plasmonic metamaterial can reach a value around 400, demonstrating an ultrahigh enhancement of the sensing performance. Third, a reliable gain coefficient in the order of hundred $cm^{-1}$ is sufficient for the greatly enhanced sensing performance, not only because the underlying EIT mechanism is lossless, but also because the unique active mechanism in plasmonic metamaterials.[11,12,28] The possible disadvantage of this active EIT-like metamaterial sensor may be its limited sensing range, since the Gaussian-shaped gain distribution might not fully account for the EIT-like resonance shift, in circumstance of wide refractive-index changes of the surrounding medium.

This work was supported by the National Natural Science Foundation of China (No.10874081, No.10904012, and No. 60990320), the National Key Projects for Basic Researches of China (No. 2010CB630703), and the Research Fund for the Doctoral Program of Higher Education of China (No. 20090092120031).

**Figure captions**

FIG. 1. (Color online) The schematic illustration of the metallic metamaterial. The EIT-like phenomenon happens when the upper bar is not at the half-length plane of the bottom double bars (i.e., when the structure is asymmetric with respect to the polarized *E* direction, corresponding to the symmetry-broken shift $s \neq 0$).

FIG. 2. (Color online) The EIT-like transparency window in dependence on the asymmetric shift *s* of the perpendicularly stacked bar, for gain coefficient $\alpha = 0 \text{ cm}^{-1}$.

FIG. 3. (Color online) The enhanced transparency windows at different gain coefficients, calculated at asymmetric shift $s = 20 \text{ nm}$.

FIG. 4. (Color online) The sensing performance for the active EIT-like transparency window. (a) The resonance shift of the EIT-like peak on the refractive-index fluctuation of surrounding medium, calculated at gain coefficient $\alpha = 850 \text{ cm}^{-1}$. (b) The FOM of the active EIT-like transparency peak in dependence on the gain coefficient, calculated at the refractive index of surrounding medium $n = 1$. The red curve provides a Lorentzian fitting.

FIG. 1. Dong et al.

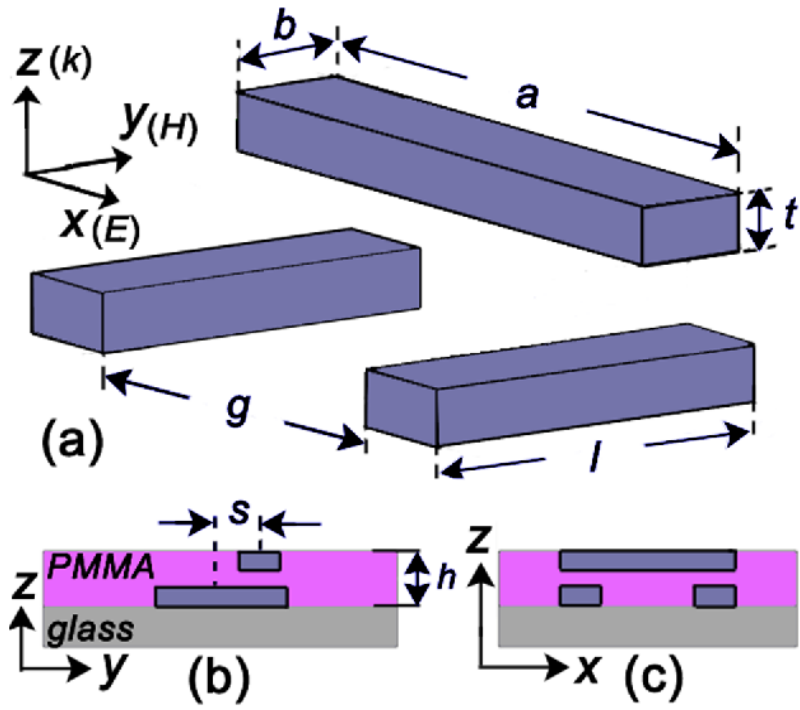

FIG. 2. Dong et al.

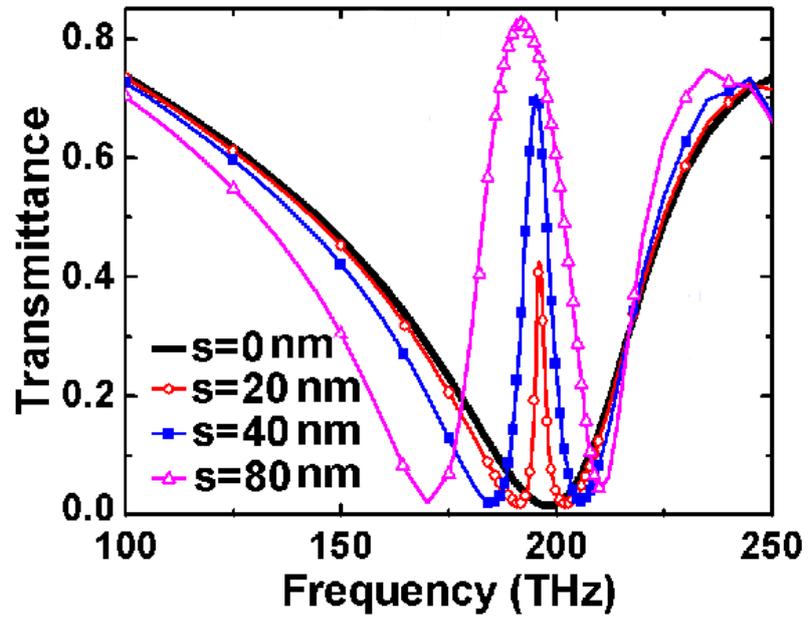

FIG. 3. Dong et al.

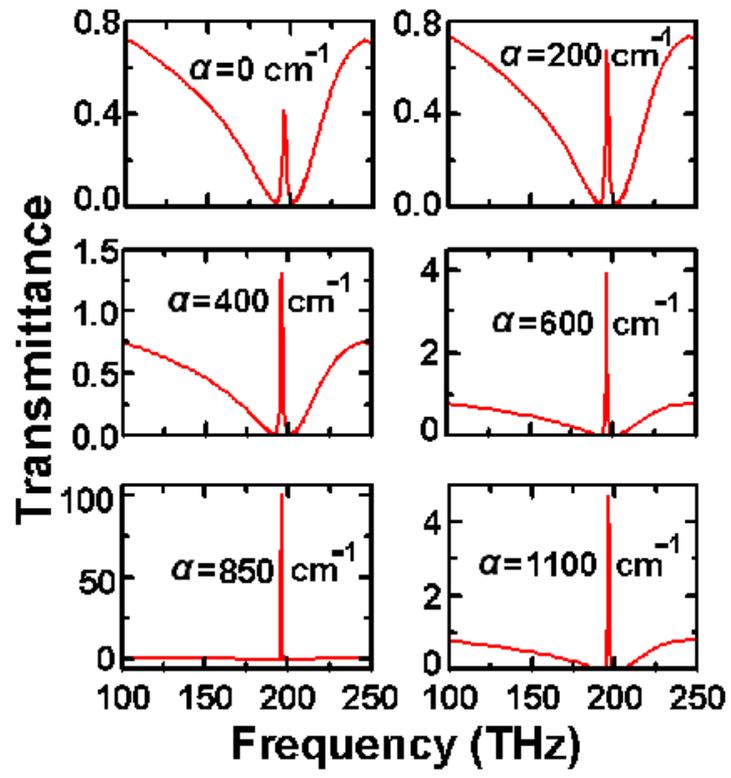

FIG. 4. Dong et al.

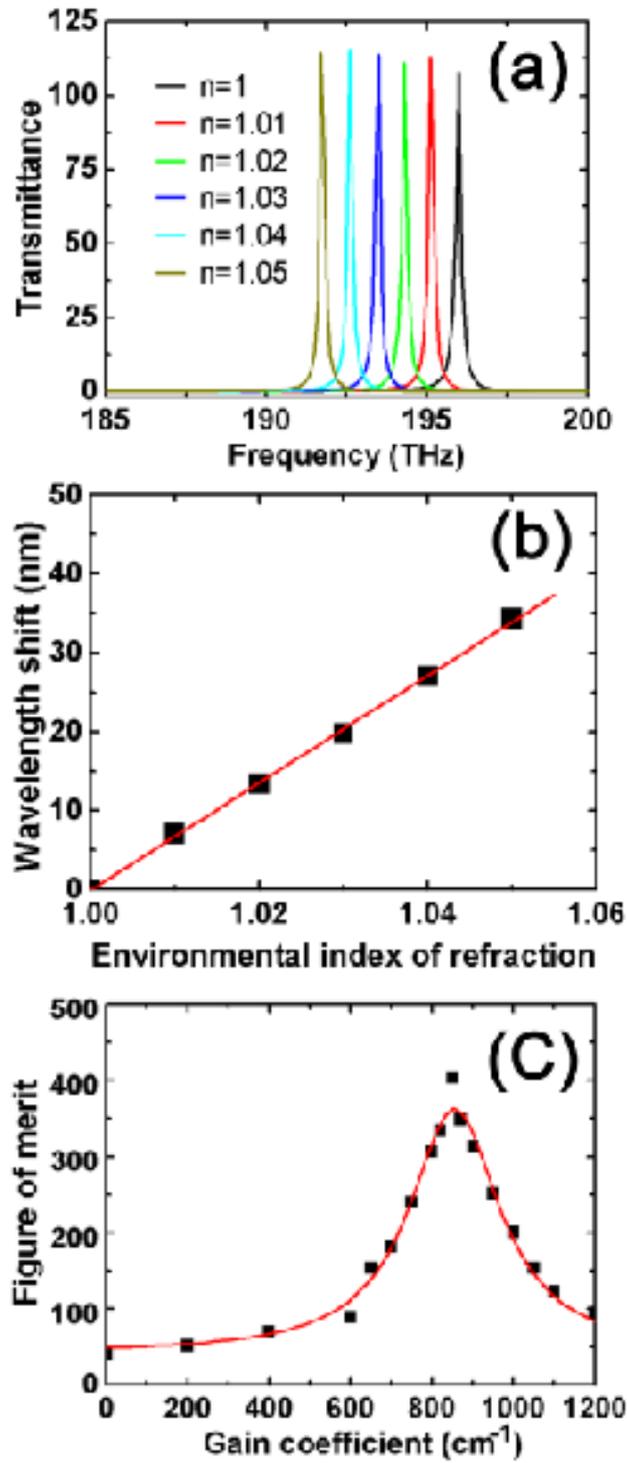